\documentclass[10pt,journal]{IEEEtran}
\IEEEoverridecommandlockouts

\usepackage{graphicx}
\usepackage{amsmath,amsthm,amsfonts,amssymb}
\usepackage{cite,hyperref}
\usepackage{bm}
\usepackage{bbm}
\usepackage{url}
\usepackage{array}
\usepackage{color,soul}
\usepackage{multirow}
\usepackage{booktabs}
\usepackage[table,xcdraw]{xcolor}
\usepackage{enumitem}
\usepackage{subcaption}

\usepackage[english]{babel}
\usepackage{algorithm}
\usepackage{algorithmic}
\usepackage{setspace}
\usepackage[english]{babel}

\theoremstyle{plain}

\newtheorem{rem}{Remark}
\newtheorem{sty1}{Theorem}
\newtheorem{defi}[sty1]{Definition}

\allowdisplaybreaks[4]

\begin{document}
\title{Deep Variable-Length Feedback Codes}

\author{Yu~Ding and
        Yulin~Shao
\thanks{The authors are with the Department of Electrical and Electronic Engineering, The University of Hong Kong, Hong Kong S.A.R. (e-mails: \url{yuding.um@gmail.com}, \url{ylshao@hku.hk}). 
}
\thanks{The source codes and well-trained DeepVLF models are available online at \url{https://github.com/hku-icl/DeepVLF}.}
}

\maketitle

\begin{abstract}
Deep learning has enabled significant advances in feedback-based channel coding, yet existing learned schemes remain fundamentally limited: they employ fixed block lengths, suffer degraded performance at high rates, and cannot fully exploit the adaptive potential of feedback. This paper introduces Deep Variable-Length Feedback (DeepVLF) coding, a flexible coding framework that dynamically adjusts transmission length via learned feedback. We propose two complementary architectures: DeepVLF-R, where termination is receiver-driven, and DeepVLF-T, where the transmitter controls termination. Both architectures leverage bit-group partitioning and transformer-based encoder-decoder networks to enable fine-grained rate adaptation in response to feedback.
Evaluations over AWGN and 5G-NR fading channels demonstrate that DeepVLF substantially outperforms state-of-the-art learned feedback codes. It achieves the same block error rate with $20$-$55\%$ fewer channel uses and lowers error floors by orders of magnitude, particularly in high-rate regimes. Encoding dynamics analysis further reveals that the models autonomously learn a two-phase strategy analogous to classical Schalkwijk-Kailath coding: an initial information-carrying phase followed by a noise-cancellation refinement phase. This emergent behavior underscores the interpretability and information-theoretic alignment of the learned codes.
\end{abstract}


\section{Introduction}
Channel coding in the presence of feedback has been a pivotal research area in information and coding theory for decades \cite{shannon1956zero,schalkwijk1966coding1,kim2011error,polyanskiy2011feedback,ben2017interactive,lai2025variable}. The classical feedback channel model, established by Shannon in 1956 \cite{shannon1956zero}, involves data transmission from a transmitter to a receiver over a memoryless noisy channel, with the receiver providing real-time feedback to aid the forward channel coding. Although feedback does not increase the forward channel capacity, it significantly enhances communication reliability in the finite block length regime. A seminal example is the Schalkwijk-Kailath (SK) scheme \cite{schalkwijk1966coding1}, which achieves a double exponential decay of the block error rate with respect to the block length, underscoring the profound impact of feedback on error performance. Beyond theoretical interest, feedback channel coding is essential to realizing Rich-ARQ systems \cite{RichARQ}, where feedback carries not just a single acknowledgement bit, but rich, soft decoding information, enabling far more efficient retransmission and adaptation strategies than conventional 1-bit ARQ.

Designing feedback channel codes, however, is considerably more intricate than designing forward codes. In forward channel coding, the encoding strategy is fixed and depends only on the information bits. In contrast, feedback codes must dynamically adapt to the receiver's state based on noisy, sequential feedback \cite{polyanskiy2011feedback,ben2017interactive}, requiring real-time, interactive decision-making. This complexity has historically limited the development of efficient and implementable feedback coding schemes.

In recent years, deep learning (DL) has emerged as a promising approach for designing sophisticated communication systems \cite{o2017introduction,shao2024theory}, including feedback channel codes \cite{kim2018deepcode,ozfatura2022all,shao2024deep}. DL models can learn complex encoding and decoding structures directly from data, capturing intricate patterns and dependencies inherent in the communication and feedback processes. A pioneering effort in this domain is DeepCode \cite{kim2018deepcode}, where the authors proposed a framework utilizing bit-by-bit passive feedback: the receiver feeds back raw, noise-contaminated symbols without processing them. In DeepCode, recurrent neural networks (RNNs) \cite{medsker2001recurrent} at both the encoder and decoder create and exploit correlations among source bits and receiver feedback, improving decoding performance.

AttentionCode \cite{shao2023attentioncode} introduced attention mechanism-based deep neural networks (DNNs) \cite{vaswani2017attention} to replace RNNs, enabling the creation of longer-range correlations among source bits and feedback. In conjunction with a bit alignment mechanism, AttentionCode achieves significantly lower block error rates (BLERs) compared to DeepCode. To enhance error-correction performance, the Deep Extended Feedback Code (DEFC)  \cite{safavi2021deep} processes feedback over extended time windows and supports higher-order modulation. The state-of-the-art was improved significantly with the introduction of Generalized Block Attention Feedback (GBAF) codes \cite{ozfatura2022all}, which partition message bits into small groups and perform group-level encoding and decoding using a sequence-to-sequence transformer architecture. This strategy efficiently reduces the input length to the DNN, thereby simplifying the learning process. Subsequent work \cite{ozfatura2023feedback} extended GBAF to active feedback scenarios, where the receiver processes the received signal before feeding it back to the transmitter. Additionally, LightCode \cite{ankireddy2024lightcode} was proposed as a lightweight coding architecture that achieves performance comparable to GBAF but with significantly fewer learning parameters. 

The scope of DL-based feedback coding has also expanded beyond single-user point-to-point channels to encompass a broader range of communication scenarios. For example, in multi-user environments, learned nonlinear codes have been developed for Gaussian broadcast channels with feedback \cite{malayter2025deep,zhou2025learned}, demonstrating enhanced robustness to feedback noise compared to conventional linear schemes. In the domain of physical-layer security, modular learned coding schemes that exploit feedback have been introduced for wiretap channels \cite{zhou2025feedback}, achieving positive secrecy rates even under reversely degraded conditions.

Despite these advancements, existing DL-aided feedback codes are predominantly fixed-length, which limits their adaptability and prevents them from fully exploiting the potential of rich feedback. 
Moreover, these codes perform well in the low code rate regime (typically lower than $1/3$), but their performance deteriorates significantly as the code rate increases, where fewer channel uses are available for transmission.
To bridge this gap, this paper introduces Deep Variable-Length Feedback (DeepVLF) coding, a new paradigm that dynamically adjusts transmission length via learned feedback.
Our core contributions are threefold:
\begin{itemize}[leftmargin=0.5cm]
    \item We propose the first deep learning framework for variable-length feedback coding, introducing two new architectures: DeepVLF-R (receiver-termination) and DeepVLF-T (transmitter-termination). By partitioning messages into bit groups and making real-time termination decisions, these codes fundamentally break the fixed-length constraint, enabling adaptive and fine-grained rate control that fully leverages feedback.
    \item DeepVLF sets a new state-of-the-art, particularly in the high-rate regime where prior learned codes falter. Extensive experiments over AWGN and 5G-NR fading channels demonstrate that our approaches achieve target reliability with over $20\%$ to $55\%$ fewer channel uses compared to the best existing fixed-length feedback codes, while pushing error floors orders of magnitude lower.
    \item Through encoding dynamics analysis, we reveal that DeepVLF models autonomously learn the iconic two-phase strategy of classical SK coding: transmitting information first, then refining through noise cancellation. This provides crucial interpretability, demonstrating that data-driven methods can inherently converge to information-theoretically optimal principles.
\end{itemize}

\section{System Model}\label{sec:II}
We consider the classical setting of feedback communication, wherein a transmitter (user A) sends information to a receiver (user B) aided by a real-time feedback link from B to A.
In practice, user A typically corresponds to a user equipment (UE), while user B represents a base station (BS).
The scenario under study is therefore uplink data transmission with downlink feedback.
In the literature, the forward channel is often modeled as an additive white Gaussian noise (AWGN) channel, whereas the feedback channel is often assumed noiseless to enable the full utilization of feedback capabilities \cite{schalkwijk1966coding1}. The assumption of error-free feedback can be justified by the BS's higher downlink transmission power and its directional transmission capabilities with MIMO beamforming.

\begin{figure}[t]
    \centering
    \includegraphics[width=1\linewidth]{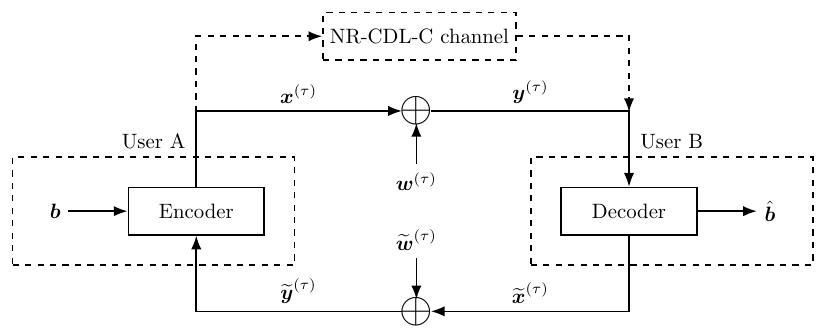}
    \caption{System model of feedback channel coding.}
    \label{fig:system_model}
\end{figure}

In this paper, we initially model both forward and feedback channels as AWGN channels to delineate our coding schemes. Later, in Section \ref{sec:V}, we will extend the channel model to a clustered delay line (CDL) fading model, as defined in 5G New Radio (3GPP TR38.901 NR-CDL-C \cite{3gppTR38901}), to emulate real-world uplink and downlink conditions and to evaluate the performance of the proposed DeepVLF codes.

Let $\bm{b}\in\{0,1\}^{K\times1}$ denote the $K$ information bits to be communicated from user A to B, as depicted in Fig. \ref{fig:system_model}. Leveraging real-time feedback, the communications take place in an interactive fashion, and can be organized into multiple rounds, each consisting of a forward transmission from A to B, followed by a feedback response from B to A. This cyclical interaction allows for continual adjustments to the coding strategy of user A based on the feedback, gradually increasing the likelihood of successful transmission.

We use the superscript $\tau\in\mathbb{N}^*$ to denote the communication round. In round $\tau$, the number of parity symbols transmitted over the forward and feedback links are denoted by $n^{(\tau)}$ and $\widetilde{n}^{(\tau)}$, respectively; both are integers and may vary across rounds.
During the forward transmission phase of round $\tau$, user A constructs a packet of parity symbols $\bm{x}^{(\tau)}\in\mathbb{R}^{n^{(\tau)}}$ utilizing both the information bits $\bm{b}$ and the feedback accumulated from previous rounds. This packet is then transmitted over forward channel. The received signal at user B is
\begin{equation}\label{eq:y}
\bm{y}^{(\tau)}=\bm{x}^{(\tau)}+\bm{w}^{(\tau)}.
\end{equation}
For any round $\tau$, the parity symbols are subject to a power constraint:
\begin{equation*}
    \mathbb{E}\left[\frac{1}{n^{(\tau)}} \langle \bm{x}^{(\tau)}, \bm{x}^{(\tau)} \rangle \right] \leq 1.
\end{equation*}

Denote by $\bm{\widetilde{x}}^{(\tau)}\in \mathbb{R}^{\widetilde{n}^{(\tau)}}$ the symbol vector to be fed back from user B, the signal received by user A is given by
\begin{equation}\label{eq:ytilde}
\bm{\widetilde{y}}^{(\tau)}=\bm{\widetilde{x}}^{(\tau)}+\bm{\widetilde{w}}^{(\tau)},
\end{equation}
where $\mathbb{E}\left[\frac{1}{\widetilde{n}^{(\tau)}} \langle \bm{\widetilde{x}}^{(\tau)}, \bm{\widetilde{x}}^{(\tau)} \rangle \right] \leq 1$.

In \eqref{eq:y} and \eqref{eq:ytilde}, $\bm{w}^{(\tau)}$ and $\bm{\widetilde{w}}^{(\tau)}$ are noise vectors consisting of independent and identically distributed (i.i.d.) zero-mean Gaussian random variables with variances $\sigma_f^2$ and $\sigma_b^2$, respectively. The case that $\sigma_b^2=0$ is termed noiseless feedback. We define the SNRs of the forward and feedback channels as $\eta_f \triangleq 10\text{log}_{10}(1/\sigma_f^2)$ and $\eta_b \triangleq 10\text{log}_{10}(1/\sigma_b^2)$, respectively. 

Depending on whether user B processes its received signal before feedback, we distinguish between \textit{passive} and \textit{active} feedback. 
Feedback is termed {passive} if $\bm{\widetilde{x}}^{(\tau)}$ is simply a scaled version of $\bm{y}^{(\tau)}$ constrained by the power limit, implying no processing at B. Conversely, feedback is termed {active} if $\bm{\widetilde{x}}^{(\tau)}$ is generated by applying some processing to $\bm{y}^{(\tau)}$.

The availability of feedback enables user A to glean insight into the noise realizations and the decoding progress at user B.
With this knowledge, user A can refine the encoding strategy in future rounds to minimize the impact of noise and improve the decoding performance. Consequently, the fundamental challenge in feedback channel coding lies in the strategic design of both the encoder and decoder to utilize this feedback effectively, enhancing the overall reliability and efficiency of the communication process.

VLF codes can be broadly categorized by who decides when to stop transmitting. Following the framework of Polyanskiy et al. \cite{polyanskiy2011feedback}, we have VLF with receiver-termination (VLF-R) and VLF with transmitter-termination (VLF-T). The distinction is that in VLF-R, the receiver decides when to terminate transmission, whereas in VLF-T, the transmitter makes that decision. Correspondingly, the DeepVLF codes proposed in this paper come in two versions: DeepVLF-R and DeepVLF-T. Before delving into their design, we formalize these concepts by adapting the definitions from \cite{polyanskiy2011feedback} to suit our framework.

\begin{defi}[VLF-R]\label{def:VLF}
    An ($\ell, K, \xi^*$) VLF-R code, where $\ell$ is a positive real, $K$ is a positive integer and $0\leq\xi^*\leq1$, is defined by
\begin{enumerate}[leftmargin=0.5cm]
  \item A sequence of encoders: ${f}_\tau:\{0,1\}^K\times \mathbb{R}^{\sum_{i=1}^{\tau-1}{{n}^{(i)}}}\times\mathbb{R}^{\sum_{i=1}^{\tau-1}{\widetilde{n}^{(i)}}}\rightarrow\mathbb{R}^{n^{(\tau)}}$, $\tau\geq1$, with
  \begin{equation*}
      \bm{x}^{(\tau)}={f}_\tau\left(\bm{b},\bm{{x}}^{(1):(\tau-1)},\bm{\widetilde{y}}^{(1):(\tau-1)}\right).
  \end{equation*}
  \item A sequence of decoders: $g_\tau:\mathbb{R}^{\sum_{i=1}^{\tau}{n^{(i)}}}\rightarrow\{0,1\}^K$ producing an estimate $\hat{\bm{b}}$ after round $\tau$;
  \item A stopping time $\tau^*$, which is a positive integer-valued random variable adapted to the termination condition $\mathcal{G}_\tau^*(\widetilde{\bm{y}}^{(1)}, \dots, \widetilde{\bm{y}}^{(\tau)})$ and satisfies $\mathbb{E}[\tau^*] \leq \ell$. The final decision is made at time $\tau^*$ as
  \begin{equation*}
      \hat{\bm{b}} = g_{\tau^*}\left(\bm{y}^{(1)},...,\bm{y}^{(\tau^*)}\right);
  \end{equation*}
  \item An upper bound on the block error rate (BLER):
  \begin{equation*}
      \mathbb{P}(\hat{\bm{b}}\neq\bm{b}) \leq \xi^*.
  \end{equation*}
\end{enumerate}
\end{defi}

In item 3) of Definition \ref{def:VLF}, the termination condition $\mathcal{G}_\tau^*(\bm{\widetilde{y}}^{(1)},...,\bm{\widetilde{y}}^{(\tau)})$ is evaluated at the receiver. This scheme is called receiver termination: the decision to end communication is made solely by user B based on its received signals. Specifically, user B terminates the process when it believes all information bits have been successfully decoded, and it then sends a termination signal to user A.

\begin{defi}[VLF-T]
    An ($\ell, K, \xi^*$) VLF-T code is defined analogously to a VLF-R code, except that condition 3) in Definition \ref{def:VLF} is replaced by
    \label{def:VLFT}
\begin{enumerate}
  \item[3')] A stopping time $\tau^*$, adapted to the termination condition $\mathcal{G}_\tau^*(\bm{b}, \bm{y}^{(1)}, \dots, \bm{y}^{(\tau)}, \widetilde{\bm{y}}^{(1)}, \dots, \widetilde{\bm{y}}^{(\tau)})$ and satisfies $\mathbb{E}[\tau^*] \leq \ell$. The final decision is
  \begin{equation*}
      \hat{\bm{b}} = g_{\tau^*}\left(\bm{y}^{(1)},...,\bm{y}^{(\tau^*)}\right).
  \end{equation*}
\end{enumerate}
\end{defi}

Unlike receiver termination, transmitter termination shifts the responsibility to user A. Since user A possesses both the true information bits $\bm{b}$ and (through feedback) an understanding of user B's estimated bits $\hat{\bm{b}}$, it can determine when decoding is complete. Once user A decides to terminate, it sends a termination signal to user B to conclude the session.

Compared to VLF-R, VLF-T incorporates a dedicated, use-once termination symbol (typically transmitted over a reliable control channel and assumed error-free) to ensure robust communication closure. This mechanism allows the transmitter to leverage feedback for a more informed and reliable termination decision, thereby reducing the risk of premature or delayed termination and significantly improving communication efficiency.

Having established the foundational framework of VLF-R and VLF-T codes, we now proceed to their neural implementation. The core challenge lies in: how can we realize the intricate encoders ${f}_\tau$ and decoders ${g}_\tau$, along with the termination policy $\mathcal{G}_\tau^*$, using DL models capable of learning from data? The following two sections answer this question by introducing DeepVLF-R (Section~\ref{sec:III}) and DeepVLF-T (Section~\ref{sec:IV}), respectively.

\section{DeepVLF-R}\label{sec:III}
The key contribution of our work lies in the data-driven realization of VLF codes. In this section, we outline the architecture and realization of the DeepVLF-R code. We will detail how neural architectures are leveraged to fulfill the roles defined in VLF codes, highlighting the key innovations that differentiate it from prior fixed-length designs.

\subsection{DeepVLF-R: Key Elements}
At the core of the DeepVLF-R architecture are several fundamental elements that guide its design.

\begin{defi}[Bit group]
    The encoding and decoding of our approaches are organized around groups of bits instead of individual bits. Given a vector of message bits $\bm{b}$, we first divide $\bm{b}$ into $Q$ equal-size groups, with each group containing $m \triangleq K/Q$ bits:
    \begin{equation*}
        \bm{b}=\left[\bm{b}_1^\top,\bm{b}_2^\top,\bm{b}_3^\top,\cdots,\bm{b}_Q^\top\right]^\top,
    \end{equation*}
    where each group $\bm{b}_q \in \{0,1\}^m$.
 \label{def:bit_group}
\end{defi}
The total number of possible patterns per group is $|\{0,1\}^m|=2^m$. This grouping approach is analogous to quadrature amplitude modulation (QAM), where symbols are formed by grouping multiple bits. This approach offers two key benefits.

First, it allows us to strike a balance between estimation complexity and decision reliability. Estimating all $K$ bits jointly would require the decoder to model a probability distribution over an output space of size $2^K$, which leads to a dense decision partition in the output domain for large $K$. Such densely packed decision boundaries make the learning process highly sensitive to small estimation errors, resulting in unstable decisions during inference.
Conversely, estimating each bit independently simplifies the per-decision process but necessitates $K$ sequential inferences, causing the overall system error rate to accumulate undesirably. By grouping bits into smaller units of size $m$, we reduce the per-inference output space to a tractable size of $2^m$, while also cutting the required number of inferences from $K$ to $Q=K/m$. This balances per-round complexity with overall estimation efficiency.

Second, grouping enables parameter sharing across bit groups, substantially improving model efficiency \cite{ozfatura2022all}. Since each group has the same structure (i.e., 
$m$ bits), the same feature extraction network can be applied to all groups, akin to weight sharing in convolutional neural networks. This reuse dramatically reduces the total number of learnable parameters, leading to a more compact and generalizable model architecture without sacrificing representational capacity.

Following this, user B employs a belief matrix to make predictions for the bit groups.
\begin{defi}[Belief matrix]\label{def:belief_matrix}
    In DeepVLF-R, user B formulates a belief matrix $\bm{P}$ to represent each transmitted bit group:
    \begin{equation*}
        \bm{P}=\left[\bm{p}_1,\bm{p}_2,\bm{p}_3,\cdots,\bm{p}_Q\right],
    \end{equation*}
    where column vector $\bm{p}_q$ corresponds to bit group $\bm{b}_q$. 
    Specifically, $\bm{p}_q$ is a $2^m$-dimensional probability distribution, with the $j$-th element indicating the likelihood that user B interprets the transmitted bit group $\bm{b}_q$ as the $j$-th element of set $\{0,1\}^m$.

\end{defi}


Contrary to existing code designs, where decoding is performed only after the complete sequence of transmissions is received by the decoder \cite{kim2018deepcode,safavi2021deep,shao2023attentioncode,ozfatura2022all,ozfatura2023feedback,ankireddy2024lightcode,shao2024deep,malayter2025deep,zhou2025feedback}, DeepVLF-R employs threshold decoding and iteratively updates its belief at the end of each communication round.

\begin{defi}[Threshold decoding]
    \label{def:threshold_decoding}
    In DeepVLF-R, user B utilizes a threshold decoding strategy to decode each bit group individually. Specifically, a bit group $\bm{b}_q$ is considered successfully decoded by user B if any element of the belief vector $\bm{p}_q^{(\tau)}$ surpasses a predefined threshold $\gamma$, i.e.,
    \begin{equation*}
        \|\bm{p}_q^{(\tau)}\|_\infty\geq\gamma,
    \end{equation*}
    where $\|\cdot\|_\infty$ denotes the infinity norm, which evaluates the maximum absolute value among the elements of a vector.
    
    When all the bit groups are deemed successfully decoded, the whole communication process is terminated. The total number of communication rounds $\tau^*$ is defined as
    \begin{equation*}
        \tau^* \triangleq \min\{\tau \mid \|{\bm{p}}_q^{(\tau)}\|_\infty \geq \gamma, q=1,2,\cdots, Q \}.
    \end{equation*}
\end{defi}

DeepVLF-R iteratively updates the belief matrix ${\bm{P}}^{(\tau)}$ at the end of each communication round. There are two types of feedback information that user B provides to user A:
\begin{itemize}[leftmargin=0.5cm]
    \item A scaled version of received symbols, i.e., $\bm{\widetilde{x}}^{(\tau)}=C_s\bm{y}^{(\tau)}$, where $C_s \triangleq \sqrt{1/(1+\sigma_b^2)}$ is a scaling factor to meet the feedback power constraint.
    \item The indices of the bit groups that have been successfully decoded within the last communication round.
\end{itemize}

Compared with the scaled received symbols $\bm{\widetilde{x}}^{(\tau)}$, the indices of the successfully decoded bit groups constitute only a very small overhead, typically just a few bits. Therefore, we assume these indices can be fed back reliably from the receiver to the transmitter via dedicated control channels.

\subsection{DeepVLF-R: Encoder and Decoder} 
The encoding process of DeepVLF-R is iterative, unfolding across several rounds. Each round leverages feedback to refine the generation of parity symbols, thereby enhancing the robustness and efficiency of the transmission. This section details the mechanics of how feedback is incorporated within each round and the subsequent generation of parity symbols. 

\subsubsection{Encoder} 
\begin{figure}[t] 
    \centering
    \includegraphics[width=0.35\textwidth]{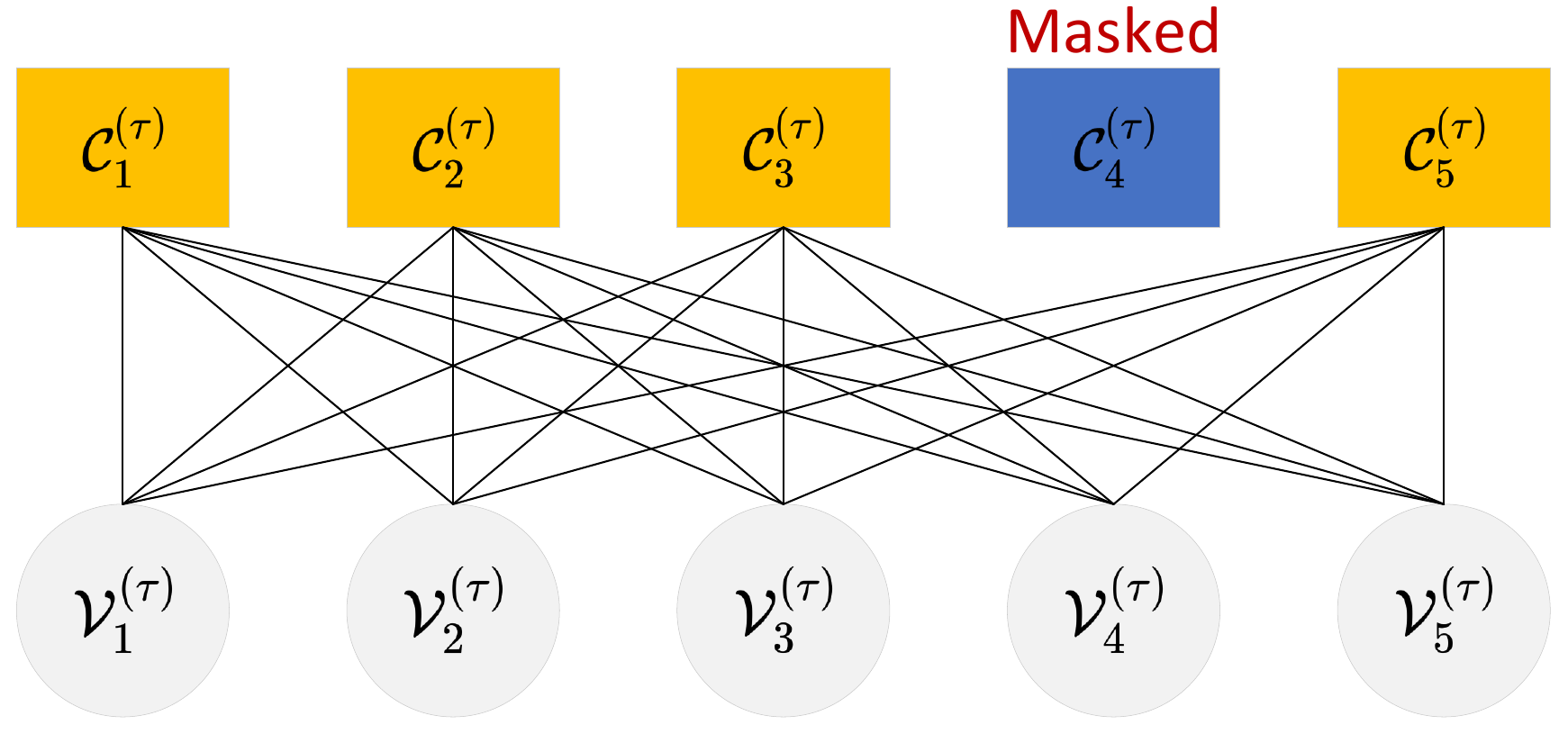} 
    \caption{The generation of parity symbols in each communication round can be represented by a bipartite graph.}
    \label{fig:sample_of_VLF}
\end{figure}

During each communication round $\tau$, the generation of parity symbols $x^{\tau}$ can be represented by a bipartite graph with variable nodes $\mathcal{V}^{(\tau)}_i, i=1,2,\cdots, Q$, check nodes $\mathcal{C}^{(\tau)}_i, i=1,2,\cdots, Q$, and a set of edges $\mathcal{E}^{(\tau)}$, as shown in Fig. \ref{fig:sample_of_VLF}.

\begin{itemize}[leftmargin=0.5cm]
    \item Each $q$-th variable node and check node is uniquely associated with the $q$-th bit group, designed specifically to cater to the encoding and decoding needs of that group. This ensures targeted processing of each bit group throughout the communication rounds.
    \item Each check node contains a parity symbol intended for transmission to user B. As stated in Definition \ref{def:threshold_decoding}, DeepVLF-R actively decodes the bit groups in all interactions. If the $q$-th bit group $\bm{b}_q$ is decoded successfully during communication round $\tau$, user A will be informed of this through feedback, and the parity symbol in $\mathcal{C}^{(\tau)'}_q, \tau'>\tau$, will no longer be transmitted to user B. In each communication round $\tau$, the number of transmitted symbols is equal to the number of bit groups that have not been decoded, i.e.,
    \begin{equation*}
        \label{equ:VLF_n_tau}
        n^{(\tau)}=\left|\{\bm{b}_q \mid \| \bm{p}_q^{(\tau)}\|_\infty<\gamma \}\right|.
    \end{equation*}
    \item The variable nodes contain the information to generate the check nodes. Initially, in the first round, we have $\mathcal{V}^{(1)}_q=\{\bm{b}_q\}$. With each subsequent communication round, $\mathcal{V}^{(\tau)}_q$ is updated to incorporate additional feedback from user B, enriching the information used for future encoding decisions and enhancing the system's adaptability to channel noise realizations.
\end{itemize}

The check nodes and variable nodes are interconnected through a set of edges. Let $e_{ij}$ represent a possible edge linking the $i$-th variable node with the $j$-th check node. During communication round $\tau$ , the set of active edges is denoted by $\mathcal{E}^{(\tau)}$. The configuration of $\mathcal{E}^{(\tau)}$ is dynamically adjusted based on the decoding status of bit groups as follows:
\begin{itemize}[leftmargin=0.5cm]
    \item Initially, when no bit group has been decoded, all possible connections are active, thus $\mathcal{E}^{(1)} = \{e_{ij}, \forall i, j\}$.
    \item If only one bit group $\bm{b}_q$ was successfully decoded in the previous round $\tau-1$, the edges connecting to the corresponding check node q are deactivated in the current round. Therefore, $\mathcal{E}^{(\tau)}=\mathcal{E}^{(\tau-1)}\setminus\{e_{iq},\forall i\}$
    \item Each edge $e_{ij} \in \mathcal{E}^{(\tau)}$ is associated with a weight $\rho_{ij}$. For each $\mathcal{C}_j$, we have $\sum_{i=1}^Q\rho_{ij}=1$.
\end{itemize}

Given the bipartite graph, the relationship between the variable and check nodes can be written as:
\begin{equation}
    \label{equ:bipratite_enc}
    \mathcal{C}_j^{(\tau)}=f^{(c)}\left(\sum_{i=1}^Q\rho_{ij}f^{(v)}(\mathcal{V}_i^{(\tau)})\right),
\end{equation}
where the encoding functions $f^{(c)}$ and $f^{(v)}$ are realized by DNNs. The coefficients $\rho_{ij}$ are learnable parameters, which are optimized to effectively determine the influence of each variable node on the corresponding check node. This architecture of DNNs and their role in mapping variable nodes to check nodes is detailed further in the next subsection.

The updating process for variable nodes after each communication round is crucial for the overall system's adaptability and efficiency. This update is governed by an auto-regressive mechanism as described below:
\begin{equation}
    \label{equ:VLF_var_nodes}
    \mathcal{V}_q^{(\tau)}=
    \begin{cases}
    \{\bm{b}_q\}, & \mathrm{if~}\tau=1; \\
    \mathcal{V}_q^{(\tau-1)}\cup\mathcal{C}_q^{(\tau-1)}\cup\{\widetilde{{y}}_q^{(\tau-1)}\}, & \mathrm{if~}\|{\bm{p}}_q^{(\tau-1)}\|_\infty<\gamma; \\
    \mathcal{V}_q^{(\tau-1)}, & \text{otherwise.} 
    \end{cases}
\end{equation}

In the initial round ($\tau=1$), each variable node is simply initialized to the corresponding bit group. In subsequent rounds, the update depends on the decoding success: if the infinity norm of the belief vector ${\bm{p}}^{(\tau)}_q$ is below the threshold $\gamma$, indicating unsuccessful decoding, $\mathcal{V}^{(\tau)}_q$ is expanded to include the parity information from the last round and the received symbols. If the threshold is met or exceeded, further updates are unnecessary, and $\mathcal{V}^{(\tau)}_q$ retains its prior state from $\mathcal{V}^{(\tau-1)}_q$. Nevertheless, this “frozen” node will continue to contribute to the generation of other parity symbols for the remaining rounds

\subsubsection{Decoder} 
At user B, the decoder engages in decoding operations during each communication round upon receiving $\bm{y}^{(\tau)}$. The design principles of the decoder mirror those of the encoder. In the following, we succinctly outline the decoder's functionality.
 
The core functionality of the decoder within DeepVLF-R is to estimate the belief matrix $\bm{P}$. This process can also be modeled by a bipartite graph representation, consisting of variable nodes $\widetilde{\mathcal{V}}^{(\tau)}_i$, and check nodes $\widetilde{\mathcal{C}}^{(\tau)}_j$, akin to the encoder's structure in \eqref{equ:bipratite_enc}. We have
\begin{equation}
    \label{equ:bipratite_dec}
    \widetilde{\mathcal{C}}_j^{(\tau)}=\widetilde{f}^{(c)}\left(\sum_{i=1}^Q\widetilde{\rho}_{ij}\widetilde{f}^{(v)}(\widetilde{\mathcal{V}}_i^{(\tau)})\right).
\end{equation}

The variable node $\widetilde{\mathcal{V}}_i$ stores the received parity symbols from the previous communication rounds and the current belief about the $i$-th bit group ${\bm{p}}_i^{(\tau-1)}$. The check node $\widetilde{\mathcal{C}}_j$ estimates the updated belief $\bm{p}_j^{(\tau)}$ for the $j$-th bit group. The functions $\widetilde{f}^{(c)}$ and $\widetilde{f}^{(v)}$ are realized by DNNs, and $\widetilde{\rho}_{ij}$ are learnable parameters, mirroring the learnable weights $\rho_{ij}$ used in the encoder.

If $\|{\bm{p}}_i^{(\tau)}\|_\infty\geq\gamma$, the $i$-th bit group is deemed successfully decoded at communication round $\tau$. Consequently, the corresponding $\widetilde{\mathcal{V}}_i^{(\tau)}$ and $\widetilde{\mathcal{C}}_j^{(\tau)}$ are frozen and will not undergo further updates in future communication rounds, thereby preserving the state achieved at the point of successful decoding.

\begin{defi}[Code rate of DeepVLF-R]
    DeepVLF-R decodes bit groups individually across different communication rounds, thus introducing variability in the decoding timeline for each bit group. We define $\tau^*_q$ as the communication round during which the $q$-th bit group is decoded:
    \begin{equation*}
        \tau^*_q \triangleq \min\{\tau\mid \|{\bm{p}}^{(\tau)}_q\|_\infty\geq\gamma\}.
    \end{equation*}

    The overall code rate $R$ of DeepVLF-R can be written as
    \begin{equation*}
        R=\frac{K}{\sum_{q=1}^Q{\tau^*_q}}.
    \end{equation*}
\end{defi}

Given the DeepVLF-R framework described above, our primary objective is to maximize the average code rate $R$ while ensuring that the BLER $\xi= \text{Pr}(\bm{b}\neq\bm{\hat{b}})$, where $\bm{\hat{b}}$ denotes the estimator of information bits, remains below a specified target $\xi^*$ on average. This optimization is performed under an average power constraint. Formally, we have
\begin{align}
\max_{f_{c},f_{v},\widetilde{f}_{c},\widetilde{f}_{v},\{\rho_{ij},\widetilde{\rho}_{ij}\}} & \ \mathbb{E}[R] \notag \\
\text{s.t.} \quad & \mathbb{E}[\xi] \leq \xi^{*}, \notag \\
& \mathbb{E}\left[\frac{1}{\sum_{q=1}^{Q}\tau_{q}^{*}}\sum_{q=1}^{Q}\sum_{\tau=1}^{\tau_{q}^{*}}\left(x_{q}^{(\tau)}\right)^{2}\right] \leq 1.
\label{equ:DeepVLF_optimization}
\end{align}

\subsection{DeepVLF-R: Realization}
\label{Realization}
This section presents our realization of the DeepVLF-R code using a transformer-based DNN architecture \cite{vaswani2017attention}. 

\subsubsection{Parity symbol generation}

The encoding process at user A is guided by the principles outlined in  \eqref{equ:bipratite_enc}, wherein the variable nodes $\mathcal{V}_q^{(\tau)}$ are updated in each communication round $\tau$ based on the feedback received from user B following  \eqref{equ:VLF_var_nodes}.The knowledge vectors contained in the variable nodes are first projected onto a high-dimensional latent space through the feature extractor function $f^{(v)}$. Subsequently, these latent vectors are aggregated using coefficients $\rho_{ij}$ generated by self-attention layers. The aggregated latent representations are then mapped to the check nodes $\mathcal{C}_q^{(\tau)}$ via a header function $f^{(c)}$, resulting in the generation of parity symbols for transmission.

\textbf{Feature extractor} ($f^{(v)}$):
In our codes, the feature extractor $f^{(v)}$ plays a crucial role in projecting the knowledge vectors of the variable nodes into a latent space suitable for processing by self-attention. We implement $f^{(v)}$ using fully-connected layers with ReLU activation functions. 

Notably, in DeepVLF-R, the depth of the fully connected layers varies depending on the communication round $\tau$ through the introduction of a variable-depth feature extractor (VDFE), where the number of fully connected layers in $f^{(v)}$ is adjusted based on whether the current communication round $\tau$ is before or after a predefined threshold $\tau_\text{VD}$, while a shallower feature extractor suffices in the earlier rounds of communication ($\tau\leq\tau_\text{VD}$). In contrast, during the later rounds ($\tau>\tau_\text{VD}$), a deeper feature extractor is employed to model the increasing complexity and nonlinearity of the encoding function.

\begin{rem}
Our VDFE design is inspired by observations in existing DL-aided feedback codes, where the nonlinearity in the functional relationship between the input and output of the encoder increases as the communication rounds progress.
\end{rem}

\textbf{Self-attention coefficients} ($\rho_{ij}$):
The aggregation coefficients $\{\rho_{ij}\}$ determine the impact of each variable node on the check nodes during the encoding process. We employ a self-attention mechanism to dynamically generate these coefficients, leveraging the ability of attention mechanisms to capture dependencies and interactions between elements in a sequence. The effectiveness of using self-attention in feedback coding has been demonstrated in our prior works \cite{shao2023attentioncode,ozfatura2022all,ozfatura2023feedback}.

In this approach, each coefficient $\rho_{ij}$ is computed based on the inner product of the latent representations of the variable nodes:
\begin{equation*}
    \rho_{ij}=\text{softmax}\left(f^{(v)}(\mathcal{V}_i^{(\tau)})^\top f^{(v)}(\mathcal{V}_j^{(\tau)})\right),
\end{equation*}
where softmax ensures that the coefficients are non-negative and sum to one for each check node.

To realize the threshold decoding mechanism and account for the decoding status of bit groups in DeepVLF-R, we utilize masking within the self-attention layers. Specifically, we define a mask $\mathcal{M}^{(\tau)} \in \{0,1\}^Q$ for the $Q$ bit groups, representing the decoding status at communication round $\tau$:
\begin{equation*}
    \mathcal{M}_j^{(\tau)}=\begin{cases}
    1,&\text{if}\|\bm{{p}}_j^{(\tau)}\|_\infty<\gamma;\\
    0,&\text{otherwise.}\end{cases}
\end{equation*}

Thanks to the feedback channel, this mask is known to both the encoder and decoder. In the self-attention computation, we mask the check nodes whose associated bit group has been successfully decoded, as illustrated in Fig.~\ref{fig:sample_of_VLF}.

\textbf{Feedforward network ($f^{(c)}$):}
After the latent representations are aggregated using the self-attention coefficients, they are mapped to the check nodes through the header function $f^{(c)}$. We design $f^{(c)}$ as a two-layer fully-connected network with Gaussian error linear unit (GeLU) \cite{hendrycks2016gaussian} activation.

At the encoder side, the header function $f^{(c)}$ maps each aggregated latent vector to a single parity symbol $\bm{x}^{(\tau)}_i$, which is then transmitted over the forward channel to user B. At the decoder side, the corresponding header function $\widetilde{f}^{(c)}$ generates a belief vector $\bm{p}^{(\tau)}_q$ for the $q$-th bit group by adding an additional linear layer.
This setup allows the decoder to estimate a probability distribution over the possible bit group patterns, facilitating effective threshold decoding as per Definition \ref{def:threshold_decoding}.

\subsubsection{Decoding}
At user B, the decoder utilizes a similar transformer-based architecture to process the received signals and update the belief matrix $\bm{P}$. The decoder's variable nodes $\widetilde{\mathcal{V}}_i^{(\tau)}$ store the received parity symbols from previous communication rounds and the current belief about each bit group. The check nodes $\widetilde{\mathcal{C}}^{(\tau)}_j$ are responsible for generating the updated belief vectors $\bm{p}^{(\tau)}_j$, as described in \eqref{equ:bipratite_dec}. 

The decoder employs the same feature extractor and self-attention mechanism, ensuring consistency in processing and facilitating effective learning of the encoding-decoding relationship.

In DeepVLF-R, the mask vector $\mathcal{M}^{(\tau)}$ is also utilized at the decoder side to mask the decoded bit groups. Upon receiving the parity symbols $\bm{y}^{(\tau)}$, the decoder updates the belief vectors and applies the threshold decoding criterion. If $\|\bm{{p}}_j^{(\tau)}\|_\infty\geq\gamma$, the $j$-th bit group is declared successfully decoded, and its corresponding variable and check nodes are frozen in subsequent rounds.

\subsubsection{Loss function}
Given that our code designs attempt to decode the bit groups after each communication round, the training process must accommodate the dynamic nature of the decoding accuracy, which improves over successive rounds. In the following, we elaborate on the loss function designed for DeepVLF-R and discuss the strategies employed to enhance its training efficacy.

DeepVLF-R is trained in a supervised manner, aiming to minimize the discrepancy between the transmitted bit groups $\bm{b}_q$ and the estimated probabilities $\bm{p}^{(\tau)}_q$ at each round $\tau$. 
A straightforward loss function design is to sum the cross-entropy losses for every bit group across all communication rounds:
\begin{equation}
    \mathcal{L}=-\sum_{q=1}^Q\sum_{\tau=1}^{\tau_q^*}\sum_{j=1}^{2^m}\mathbb{I}_{(\bm{b}_q=\{0,1\}^m_j)}\cdot\log(p_{q,j}^{(\tau)}),
    \label{equ:loss}
\end{equation}
where $\mathbb{I}$ is the indicator function, and $\{0,1\}^m_j$ denotes the $j$-th element in the set $\{0,1\}^m$.
This loss function aims to minimize the cross-entropy between the true bit groups and the estimated probabilities at each round, encouraging the model to improve its decoding accuracy progressively.

However, due to the interactive nature of DeepVLF codes, the BLER naturally decreases as the communication rounds progress. In the initial rounds, the decoder's estimates are less accurate, resulting in higher cross-entropy losses. Consequently, when averaging the cross-entropy equally across all rounds, the loss function in \eqref{equ:loss} becomes dominated by the high losses from the early rounds. This dominance can hinder the training process, as the model may focus disproportionately on minimizing the loss in the early rounds, potentially at the expense of performance in later rounds where significant improvements occur.

To mitigate this problem and enhance the performance of DeepVLF, we introduce two advancements in our training.

First, we set a threshold $\tau^+$, and initiate decoding attempts only after communication round $\tau^+$. Communication rounds before $\tau^+$ are treated as undecodable, acknowledging that the decoder's estimates during these early rounds are not sufficiently reliable. By excluding the high-loss contributions from the initial rounds, we prevent them from dominating the training process. The value of $\tau^+$ is determined based on the Shannon limit  \cite{shannon1948mathematical}, providing a rough guideline for the minimum number of transmissions required for reliable communication given the channel conditions:
\begin{equation*}
    \tau^+\triangleq\max\left\{\mu,\left\lfloor\frac{2m}{\log_2\left(1+\eta_f\right)}\right\rfloor\right\},
    \label{def:tau_plus}
\end{equation*}
where $\eta_f$ is the SNR of the forward channel and $\mu$ is a baseline threshold for $\tau^+$, adjusted based on $\gamma$.

Second, we incorporate an exponential weighting factor into the loss function to balance the contributions of losses from different rounds. By assigning higher weights to losses from later rounds -- where the decoder estimates are more accurate -- we encourage the model to focus on improving performance in these critical stages. The final loss function for training DeepVLF is thus formulated as:
\begin{equation*}
    \mathcal{L}=-\sum_{q=1}^Q\sum_{\tau=\tau^+}^{\tau_q^*}\vartheta^{\tau-c}\sum_{j=1}^{2^m}\mathbb{I}_{(\bm{b}_q=\{0,1\}^m_j)}\cdot\log(p_{q,j}^{(\tau)}),
\end{equation*}
where $\vartheta$ is the base of the exponential weighting factor while $c$ denotes a fixed offset parameter. By carefully selecting $\vartheta$ and $c$, we can effectively balance the influence of each communication round on the total loss, ensuring that later rounds, which are more indicative of the model's ultimate performance, have a greater impact on the training process.

\section{DeepVLF-T}\label{sec:IV}

\begin{figure*}[t]
    \centering
    \includegraphics[width=0.95\textwidth]{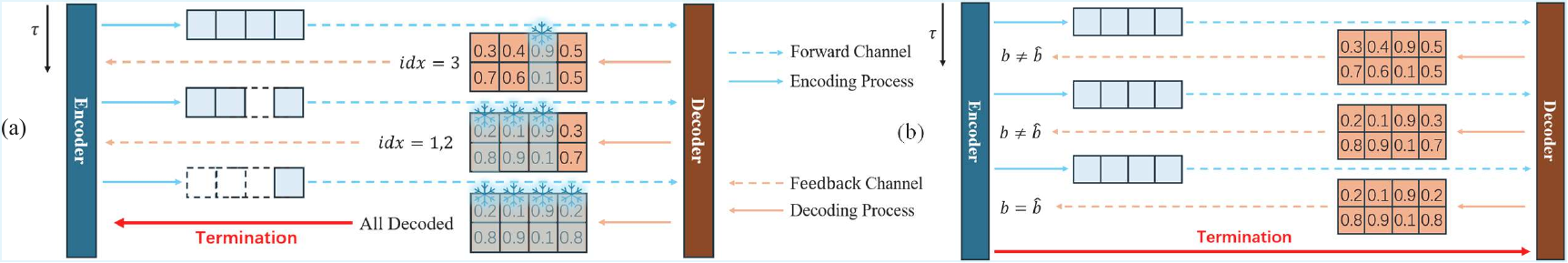} 
    \caption{The difference in coding architectures between (a) DeepVLF-R and (b) DeepVLF-T.}
    \label{fig:difference}
\end{figure*}

This section presents DeepVLF-T, the transmitter-termination counterpart to the DeepVLF-R code introduced in Section~\ref{sec:III}. While both schemes belong to the variable-length feedback coding family, they embody fundamentally different control philosophies: DeepVLF-R delegates the stopping decision entirely to the receiver, whereas DeepVLF-T entrusts this responsibility to the transmitter. This distinction leads to unique operational characteristics, advantages, and design considerations. We begin by detailing the architecture and training of DeepVLF-T, followed by an analysis of its robustness under noisy feedback. Finally, we conclude with a comparative discussion that illuminates the inherent trade-offs and complementarity between receiver-driven and transmitter-driven termination.

\subsection{Design of DeepVLF-T}
DeepVLF-T inherits the fundamental design elements of its receiver-termination counterpart, specifically, bit-group partitioning (Definition~\ref{def:bit_group}) and the belief matrix (Definition~\ref{def:belief_matrix}). The principal distinction resides in the locus of termination control and the corresponding structure of the feedback loop.

After each forward transmission round $\tau$, user B updates its belief matrix $\bm{P}^{(\tau)}$ but does not apply a final hard-decision decoding rule locally. Instead, it feeds back the following two pieces of information to user A:
\begin{itemize}[leftmargin=0.5cm]
\item A scaled version of the received symbols, $\widetilde{\bm{x}}^{(\tau)} = C_s \bm{y}^{(\tau)}$.
\item A scaled version of the current belief matrix, $C_b \bm{P}^{(\tau)}$.
\end{itemize}
Here, $C_s$ and $C_b$ are scaling factors that enforce the average power constraint on the feedback link. $C_s \triangleq \sqrt{1/(1+\sigma_b^2)}$ is identical to the scaling used in DeepVLF-R. The normalization for the belief matrix, $C_b$, is derived from the statistical properties of a probability vector. Recall that each column $\bm{p}_q^{(\tau)}$ of $\bm{P}^{(\tau)}$ is a probability distribution over $2^m$ patterns. Under a symmetric prior, each element of $\bm{p}_q^{(\tau)}$ has a variance upper-bounded by $\frac{1}{2^m} - \frac{1}{2^{2m}}$. To ensure unit average power per feedback symbol, we set
\begin{equation*}
C_b \triangleq \sqrt{\frac{1}{\frac{1}{2^m} - \frac{1}{2^{2m}}}} = \frac{2^m}{\sqrt{2^m - 1}}.
\end{equation*}

Consequently, under a noisy feedback channel with noise variance $\sigma_b^2$, the signals received by user A are
\begin{align}
\widetilde{\bm{y}}^{(\tau)} &= C_s \bm{y}^{(\tau)} + \widetilde{\bm{w}}^{(\tau)}, \label{eq:feedback_symbols_T} \\
\widetilde{\bm{P}}^{(\tau)} &= C_b \bm{P}^{(\tau)} + \widetilde{\bm{W}}^{(\tau)}, \label{eq:feedback_belief_T}
\end{align}
where $\widetilde{\bm{w}}^{(\tau)}$ and $\widetilde{\bm{W}}^{(\tau)}$ are i.i.d. Gaussian noise vectors and matrices, respectively, each with element-wise variance $\sigma_b^2$. For clarity, we first describe the core procedure assuming noiseless feedback ($\sigma_b^2 = 0$). The robust operation under noisy feedback is analyzed separately in Section~\ref{sec:robust_transmitter_termination}.

\begin{defi}[Transmitter-side termination decoding]
    Under noiseless feedback, user A receives the exact belief matrix $\bm{P}^{(\tau)}$. It forms a hard-decision estimate $\hat{\bm{b}}^{(\tau)}$ by selecting, for each bit group $q$, the pattern corresponding to the maximum entry in $\bm{p}_q^{(\tau)}$:
    \begin{equation*}
\hat{\bm{b}}_q^{(\tau)} = \mathop{\arg\max}_{j \in {1,\ldots,2^m}} \bm{P}_{q,j}^{(\tau)}.
\end{equation*}
The termination condition at user A is then defined as
\begin{equation}
\label{equ:Tx_decoding}
\mathcal{G}_\tau(\widetilde{\bm{P}}^{(\tau)}, \bm{b}) \triangleq
\begin{cases}
\mathsf{terminate}, & \text{if } \hat{\bm{b}}^{(\tau)} = \bm{b}, \\
\mathsf{continue}, & \text{otherwise}.
\end{cases}
\end{equation}
\end{defi}

When $\mathcal{G}_\tau = \mathsf{terminate}$, user A transmits a final termination signal to user B, concluding the session. The total number of communication rounds is the stopping time
\begin{equation*}
    \tau^*\triangleq\text{min}\{\tau \mid \mathcal{G}_\tau(\widetilde{\bm{P}}^{(\tau)}, \bm{b})=\mathsf{terminate} \}.
\end{equation*}

The encoding process of DeepVLF-T follows the same bipartite graph formalism as DeepVLF-R, comprising variable nodes $\mathcal{V}^{(\tau)}_q$, check nodes $\mathcal{C}^{(\tau)}_q$, and a fixed, fully-connected edge set $\mathcal{E} = \{e_{ij} \mid \forall i,j\}$. A critical difference arises from the termination logic: because the transmitter must monitor the belief state of \emph{all} bit groups to make a global termination decision, it cannot cease transmitting parity symbols for any group prematurely. Therefore, the number of transmitted symbols per round is constant and equal to the number of bit groups:
\begin{equation}
n^{(\tau)} \equiv Q, \quad \forall \tau.
\end{equation}
This also implies that the decoding timeline is synchronized across all groups: $\tau_q^* = \tau^*$ for all $q=1,\ldots,Q$.

Consequently, the variable node update rule simplifies, as there is no need to incorporate a per-group decoding mask:
\begin{equation}
    \label{equ:VLFT_var_node}
    \mathcal{V}_q^{(\tau)}=
    \begin{cases}
    \{\bm{b}_q\}, & \mathrm{if~}\tau=1; \\
    \mathcal{V}_q^{(\tau-1)}\cup\mathcal{C}_q^{(\tau-1)}\cup\{\widetilde{{y}}_q^{(\tau-1)}\}, & \text{otherwise.} 
    \end{cases}
\end{equation}
The generation of parity symbols $\bm{x}^{(\tau)}$ at the encoder and the updated belief matrix $\bm{P}^{(\tau)}$ at the decoder are performed identically to DeepVLF-R, via the transformer-based functions specified in \eqref{equ:bipratite_enc} and \eqref{equ:bipratite_dec}, utilizing separate sets of learnable attention weights $\rho_{ij}$ and $\widetilde{\rho}_{ij}$.

The overall code rate of DeepVLF-T is determined by the fixed per-round channel uses and the stopping time:
\begin{equation}
R = \frac{K}{Q \cdot \tau^*} = \frac{m}{\tau^*}.
\end{equation}
The learning objective is to optimize the neural network parameters to maximize the expected rate while satisfying the average block error probability and power constraints:
\begin{align}
\max_{f_{c},f_{v},\widetilde{f}_{c},\widetilde{f}_{v},\{\rho_{ij},\widetilde{\rho}_{ij}\}} & \ \mathbb{E}[R] \notag \\
\text{s.t.} \quad & \mathbb{E}[\xi] \leq \xi^{*}, \notag \\
& \mathbb{E}\left[\frac{1}{Q\tau^{*}}\sum_{q=1}^{Q}\sum_{\tau=1}^{\tau^{*}}\left(x_{q}^{(\tau)}\right)^{2}\right] \leq 1,
\end{align}

The training strategy, including the two-phase procedure and the customized loss function detailed in Section~\ref{Realization}, is directly applicable to DeepVLF-T. Fig.~\ref{fig:difference} presents an example to illustrate the difference between the coding architectures of DeepVLF-R and DeepVLF-T.

\subsection{DeepVLF-T Under Noisy Feedback}
\label{sec:robust_transmitter_termination}

The transmitter-termination mechanism in DeepVLF-T, as defined by \eqref{equ:Tx_decoding}, presupposes that user A has perfect access to the receiver's belief matrix $\bm{P}^{(\tau)}$. This assumption is pivotal: it allows the transmitter to accurately assess whether $\hat{\bm{b}}^{(\tau)} = \bm{b}$ with certainty. However, in practical systems, the feedback channel is inherently noisy ($\sigma_b^2>0$). Consequently, user A only observes a corrupted version $\widetilde{\bm{P}}^{(\tau)}$, as given by \eqref{eq:feedback_belief_T}. This feedback noise introduces a critical vulnerability into the termination logic.

The hard-decision rule $\hat{\bm{b}}^{(\tau)} = \mathop{\arg\max}_{j} \widetilde{\bm{P}}_{q,j}^{(\tau)}$ is highly sensitive to perturbations in $\widetilde{\bm{P}}^{(\tau)}$. Two detrimental outcomes can occur:
\begin{enumerate}
    \item \textit{False early termination}: Noise may distort $\widetilde{\bm{P}}^{(\tau)}$ such that an incorrect pattern accidentally attains the maximum and yet, by chance, matches the true $\bm{b}$ for all groups. This leads the transmitter to terminate prematurely while the receiver's actual belief $\bm{P}^{(\tau)}$ remains uncertain, resulting in a high probability of block error.
    \item \textit{Excessive communication delay}: More commonly, noise can suppress the maximum value of the true posterior, causing $\hat{\bm{b}}^{(\tau)} \neq \bm{b}$ even when $\bm{P}^{(\tau)}$ would indicate reliable decoding. This forces unnecessary additional rounds, degrading the expected code rate.
\end{enumerate}

Therefore, enhancing the robustness of the termination decision is not merely an implementation detail but a fundamental requirement for DeepVLF-T. A principled method to mitigate this sensitivity is to introduce a \textit{confidence threshold} $\widetilde{\gamma}$. The core idea is to permit termination only when the received belief vectors exhibit sufficiently high confidence, thereby providing a buffer against feedback noise. We formalize this robust termination rule as follows.

\begin{defi}[Threshold-gated transmitter termination]
Under noisy feedback, the robust termination condition at user A is:
\begin{equation}\label{eq:robust_termination_rule_v2}
\mathcal{G}'_\tau(\widetilde{\bm{P}}^{(\tau)}, \bm{b}) \triangleq 
\begin{cases} 
\mathcal{G}_\tau(\widetilde{\bm{P}}^{(\tau)}, \bm{b}), 
& \text{if } \|\widetilde{\bm{p}}_q^{(\tau)}\|_\infty \ge \widetilde{\gamma},
\\ 
&\forall q=1,\ldots,Q,
\\
\mathsf{continue}, & \text{otherwise},
\end{cases}
\end{equation}
where $\mathcal{G}_\tau(\cdot)$ is defined in \eqref{equ:Tx_decoding} and $\widetilde{\gamma} \in (1/2^m, 1)$ is a pre-defined confidence threshold. The corresponding stopping time becomes:
\begin{equation*}
\tau^* \triangleq \min \left\{ \tau \,\middle|\, 
\hat{\bm{b}}^{(\tau)}(\widetilde{\bm{P}}^{(\tau)}) = \bm{b}
\ \land\ 
\min_q \|\widetilde{\bm{p}}_q^{(\tau)}\|_\infty \ge \widetilde{\gamma}
\right\}.
\end{equation*}
\end{defi}

This mechanism acts as a gating function: the condition $\|\widetilde{\bm{p}}_q^{(\tau)}\|_\infty \ge \widetilde{\gamma}$ ensures that \emph{all} bit groups are perceived by the transmitter to have a clear, dominant candidate pattern. This significantly reduces the probability that noise alone causes the terminate condition to be met spuriously. The threshold $\widetilde{\gamma}$ parameterizes a trade-off: a higher $\widetilde{\gamma}$ demands higher confidence, increasing robustness and final reliability at the potential cost of a longer average $\mathbb{E}[\tau^*]$ (lower average rate). This trade-off can be tuned based on the feedback channel quality ($\sigma_b^2$) and the target BLER.

While the threshold-based method is intuitive and effective, it presents several considerations:
\begin{itemize}[leftmargin=0.5cm]
    \item \textit{Threshold selection}: The optimal $\widetilde{\gamma}$ depends on both $\sigma_f^2$ and $\sigma_b^2$, and may not be trivial to derive analytically. In practice, it can be treated as a hyper-parameter optimized via simulation.
    \item \textit{Suboptimality}: This heuristic rule does not necessarily minimize $\mathbb{E}[\tau^*]$ for a given target BLER under noisy feedback. It is a conservative, worst-case-inspired safeguard.
    \item \textit{Information underutilization}: The rule uses only the infinity norm of the noisy beliefs, ignoring the rich structure within $\widetilde{\bm{P}}^{(\tau)}$ and the historical feedback sequence $\widetilde{\bm{y}}^{(1:\tau)}$.
\end{itemize}

The limitations above point to a promising research direction: the termination function itself can be cast as a learned policy. Instead of a fixed rule, one could employ a lightweight neural network $f_{\text{term}}(\bm{b}, \widetilde{\bm{P}}^{(\tau)}, \widetilde{\bm{y}}^{(1:\tau)}; \bm{\theta})$ that is trained jointly with the encoder and decoder. This network could integrate all available information at the transmitter, including the known information bits $\bm{b}$ and the complete history of noisy feedback, to produce a more nuanced $\mathsf{continue}$/$\mathsf{terminate}$ decision that directly optimizes the rate-reliability trade-off under noisy feedback. Such an approach would represent a fully data-driven realization of an optimal stopping policy for VLF-T codes, moving beyond hand-designed heuristics.

\subsection{Comparative Analysis and Hybrid Design}

Having introduced both DeepVLF-R and DeepVLF-T, a critical question arises: \textit{what are the fundamental trade-offs between these two control paradigms, and can their strengths be combined}? This section provides a comparative analysis, introducing a new \textit{hybrid} framework that interpolates between the two extremes. We will show that the optimal operating point of this hybrid scheme depends critically on the reliability of the feedback channel: DeepVLF-T is favored under reliable feedback, whereas DeepVLF-R becomes advantageous as feedback noise increases.

\subsubsection{Fundamental trade-offs}
The core distinction between DeepVLF-R and DeepVLF-T lies in the locus of information and control. In DeepVLF-R, termination is distributed to the receiver, which operates under information asymmetry as it lacks knowledge of the true message $\bm{b}$. It must therefore rely on a local, per-group confidence threshold $\gamma$, making fine-grained termination decisions. This structure minimizes feedback overhead and is inherently robust to feedback noise, as the threshold provides a decision margin. However, it risks error propagation: an incorrectly set $\gamma$ can cause premature termination of a group, compromising the entire block.

In contrast, DeepVLF-T centralizes control at the transmitter, which possesses perfect knowledge of $\bm{b}$. By comparing $\bm{b}$ against the receiver's fed-back belief matrix $\widetilde{\bm{P}}^{(\tau)}$, it makes a single, global termination decision. This centralization eliminates the risk of distributed decision errors and typically yields higher reliability, as the transmitter has a complete view of the receiver's state. The cost is significantly higher feedback overhead (transmitting the belief matrix) and heightened sensitivity to noise on the feedback link, since corruption of the belief matrix directly affects the stopping decision.

\subsubsection{A unified hybrid framework}
Motivated by these complementary profiles, we propose a hybrid joint decoding framework that allows termination control to be dynamically shared between the transmitter and receiver. This framework operates as follows in each round $\tau$:
\begin{itemize}[leftmargin=0.5cm]
    \item Receiver-driven per-group termination: The receiver applies its threshold $\gamma$ to its local belief vectors $\bm{p}_q^{(\tau)}$. Any group $q$ where $\|\bm{p}_q^{(\tau)}\|_\infty \geq \gamma$ is declared locally decoded. The indices of these groups are fed back to the transmitter, and their corresponding columns of the belief matrix are frozen in subsequent rounds.
    \item Transmitter-driven global oversight: The transmitter receives the full belief matrix $\widetilde{\bm{P}}^{(\tau)}$ and the indices of receiver-decoded groups. For the remaining undecoded groups, it performs a global check: if the maximum-likelihood estimate $\hat{\bm{b}}^{(\tau)}$ derived from $\widetilde{\bm{P}}^{(\tau)}$ matches the true $\bm{b}$ for all remaining groups, it initiates a final termination.
\end{itemize}

This framework is parameterized by $\gamma$: setting $\gamma \to 1$ effectively disables receiver termination, reducing the system to DeepVLF-T; conversely, a lower $\gamma$ empowers the receiver to terminate more groups early, approximating the efficiency of DeepVLF-R while retaining the transmitter's safety check for the most uncertain bits.

To move beyond qualitative description and precisely characterize the operating point on this spectrum, we introduce a new metric: the differential code rate $R_d$ defined as
\begin{equation}\label{eq:diff_rate}
R_d = \frac{K}{\sum_{\tau=1}^{\tau^*} n^{(\tau)}} - \frac{K}{\tau^* Q}.
\end{equation}
The first term is the actual achieved rate of the hybrid scheme, which benefits from early termination of groups ($n^{(\tau)} \leq Q$). The second term is the baseline rate of a DeepVLF-T system with the same stopping round $\tau^*$, which transmits all $Q$ symbols every round. Therefore, $R_d$ directly quantifies the throughput gain attributable solely to the receiver's distributed, per-group termination decisions. A value of $R_d \approx 0$ indicates a transmitter-dominated process, while a larger $R_d$ signifies greater contribution from receiver-side early termination.

\begin{figure}
    \centering
    \includegraphics[width=1\linewidth]{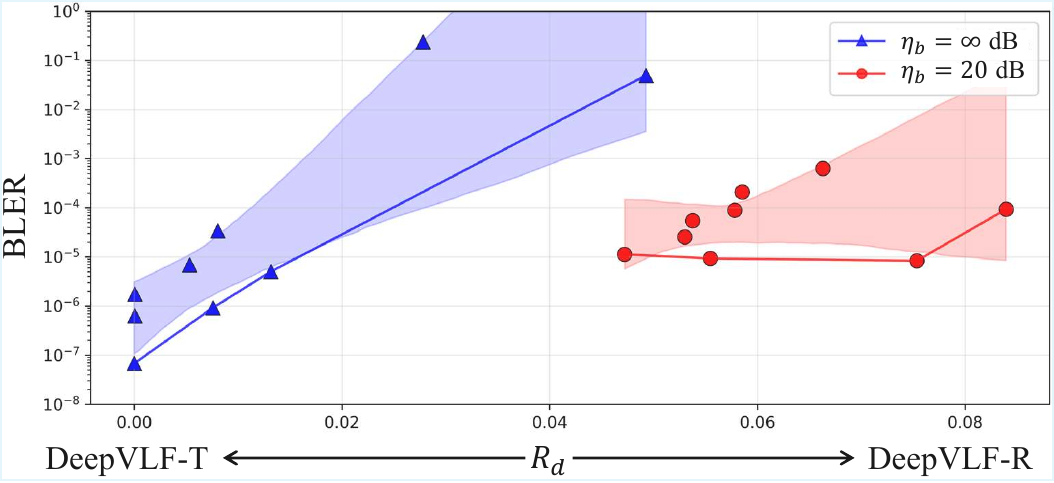}
    \caption{BLER versus the differential code rate $R_d$ for the hybrid scheme under varying receiver confidence thresholds $\gamma$. Here, $R_d$ quantifies the throughput gain from receiver-side early termination. A higher $R_d$ indicates a stronger contribution from receiver-driven per-group termination. As $R_d\to 0$, the performance of the hybrid scheme converges to that of pure DeepVLF-T, indicating that transmitter-side global verification dominates the process (this occurs notably in the ultra-reliable regime with BLER $< 10^{-4}$).}
    \label{fig:differential_code_rate_vs_PER}
\end{figure}

The empirical relationship between $R_d$ and the BLER, depicted in Fig. \ref{fig:differential_code_rate_vs_PER}, yields a key finding. In the ultra-reliable regime (BLER $< 10^{-4}$) with reliable feedback ($\eta_b=\infty \text{ dB}$), $R_d$ is negligible. This indicates that the hybrid scheme's performance converges to that of pure DeepVLF-T, as the transmitter's global verification becomes the decisive factor for achieving ultimate reliability. The potential marginal rate gain from aggressive receiver-side termination ($\gamma$ too low) is outweighed by the increased risk of undetected errors.

In contrast, under noisy feedback conditions ($\eta_b=20 \text{ dB}$), the role of transmitter-side termination becomes less pronounced due to its reliance on accurate belief feedback.
The hybrid framework can mitigate this vulnerability by lowering $\gamma$, allowing the receiver to terminate confident groups based on forward-link observations, thereby achieving a positive $R_d$ with a controlled and often modest reliability penalty.

This result leads to concrete design guidelines:
\begin{itemize}[leftmargin=0.5cm]
    \item For applications with stringent reliability targets and moderately reliable feedback, DeepVLF-T is the superior choice. Its centralized control provides the strongest guarantee of correct decoding.
    \item The hybrid framework is most valuable as a tunable system. In challenging feedback conditions, a lower $\gamma$ can be set to leverage the receiver's robustness, accepting a small reliability penalty for significant rate improvement ($R_d > 0$).
    \item DeepVLF-R represents the efficient extreme of this hybrid scheme, optimal when feedback bandwidth is minimal and the channel conditions are well-matched to a carefully calibrated $\gamma$.
\end{itemize}

Our comparative analysis clarifies that DeepVLF-T remains the benchmark for peak reliability under reliable feedback, while DeepVLF-R excels in feedback-limited scenarios.

\section{Numerical Results and Insights}\label{sec:V}
This section presents a comprehensive performance evaluation and analysis of the proposed DeepVLF framework. We benchmark DeepVLF-R and DeepVLF-T against state-of-the-art DL-based feedback codes under both controlled AWGN and realistic 5G-NR fading channels. Beyond establishing superior error-rate performance, we provide encoding dynamics analysis that elucidates the learned encoding strategies and operational principles of variable-length feedback codes.

\begin{figure*}[t]
    \centering
    \includegraphics[width=0.8\linewidth]{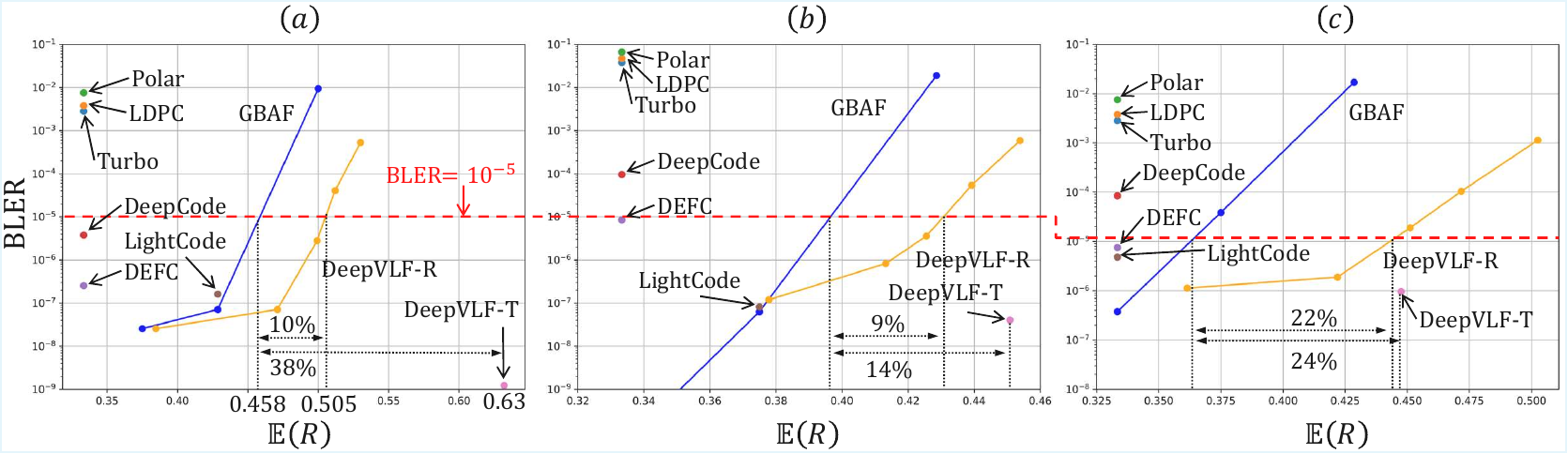}
    \caption{The BLER versus average code rate performances of DeepVLF-R and DeepVLF-T over AWGN channel: (a) $\eta_f =1 \text{dB}$, noiseless feedback; (b) $\eta_f =0 \text{dB}$, noiseless feedback; (c) $\eta_f =1 \text{dB}$, $\eta_b =20 \text{dB}$.}
    \label{fig:DeepVLF}
\end{figure*}

\subsection{Experiment Setup and Training Methodology}
Throughout our experiments, we consider a message block of $K = 48$ information bits. Following Definition \ref{def:bit_group}, the message is partitioned into $Q = 16$ groups, each containing $m = 3$ bits. This group size provides a tractable $2^m=8$-dimensional output space for the decoder's belief vector while enabling significant parameter sharing. To facilitate implementation with fixed-dimension neural networks, we define a maximum number of communication rounds, $\tau_{\text{max}}$. Input sequences are zero-padded to this length. We set $\tau_{\text{max}} = 10$ for DeepVLF-R and $\tau_{\text{max}} = 20$ for DeepVLF-T.
In model training, we employ the AdamW optimizer, a refined version of the Adam optimizer that decouples weight decay from the gradient-based updates \cite{loshchilov2017decoupled}, and a learning rate scheduling strategy where the learning rate progressively decreases as the training steps advance. To optimize the performance of our approaches across different code rates and BLER requirements, we structure the training process into two phases:
\begin{itemize}[leftmargin=0.5cm]
    \item \textbf{Phase I (Pre-training)}: The model is trained under a randomized curriculum of forward SNRs ($\eta_f$) and decoding thresholds ($\gamma$ for DeepVLF-R, $\widetilde{\gamma}$ for DeepVLF-T). This phase encourages the model to learn robust, generalizable feature representations adaptable to varying channel conditions and reliability targets.
    \item \textbf{Phase II (Fine-tuning)}: The pre-trained model is fine-tuned on specific target operational points ($\eta_f$, $\eta_b$, $\gamma$, $\widetilde{\gamma}$). This phase sharpens performance, allowing precise control over the trade-off between the average code rate $\mathbb{E}[R]$ and the BLER.
\end{itemize}

\begin{table}[t]
    \centering
    \caption{Hyperparameter settings}
    \label{tab:hyperparameter}
    \setlength{\tabcolsep}{0.4mm} 
    \begin{tabular}{cccc}
        \toprule
        \textbf{Para.} & \textbf{Descriptions}                      & \textbf{DeepVLF-R} & \textbf{DeepVLF-T} \\ 
        \midrule
        $K$                 & Bitstream length                          & 48    & 48     \\ 
        $Q$                 & Number of bit groups                      & 16    & 16     \\ 
        $m$                 & Size of a bit group                       & 3     & 3      \\ 
        $B$                 & Training batchsize                     & 8192  & 2048   \\ 
        $B_{\text{test}}$   & Test batchsize                    & 8192  & 16384  \\ 
        lr                  & Initial learning rate                     & $10^{-3}$ & $10^{-3}$     \\ 
        $\lambda$           & Weight decaying factor                    & $10^{-3}$ & $10^{-3}$     \\ 
        $\vartheta$         & Exponential factor base            & 10 & $10^{0.25}$ \\ 
        $c$                 & Exponential factor offset & 9  & 16         \\ 
        $\tau_{\text{max}}$        & Maximal interactions & 10 & 20      \\
        \bottomrule
    \end{tabular}
\end{table}

The hyper-parameters settings for the training of our codes are summarized in Table \ref{tab:hyperparameter}, unless otherwise specified.
For DeepVLF-R, the average code rate varies with the decoding threshold $\gamma$, which determines when bit groups are considered successfully decoded based on the belief matrix. In the experiments, we vary $\gamma$ from $1-10^{-3}$ to $1-10^{-7}$ to dynamically adjust its average code rate. The parameter $\mu$ defined in Section~\ref{def:tau_plus} is set to 3 for DeepVLF-T and 
\begin{equation*}
    \mu=
    \begin{cases}
    5, & \mathrm{if~}\gamma<=1-10^{-5} \\
    6, & \mathrm{if~}1-10^{-5}<\gamma<=1-10^{-6} \\
    7, & \mathrm{if~}\gamma>10^{-6}
    \end{cases}
\end{equation*}
for DeepVLF-R.

Under noisy feedback, DeepVLF-T employs a threshold-based transmitter decision decoding strategy instead of the standard transmitter decision decoding to improve robustness against feedback noise. This approach makes the transmitter's estimation of $\hat{\bm{b}}$ more resilient to noise, with its noise tolerance controlled by the threshold parameter $\gamma_{t}$. It is important to note that increasing $\gamma_{t}$ enhances robustness of the noise at the cost of a lower effective code rate, as more transmission rounds are required to satisfy the decoding criterion. This mechanism explicitly exposes the fundamental trade-off between robustness to feedback noise and transmission efficiency, which is central to transmitter-driven termination strategies.
In our experiments, $\gamma_{t}$ is set to 0.7 when the feedback channel is modeled as an AWGN channel and to 0.9 when it is a fading channel. For a noiseless feedback channel, $\gamma_{t}$ is set to 0, making threshold-based transmitter decision decoding equivalent to standard transmitter decision decoding.

\begin{figure*}[t]
    \centering
    \includegraphics[width=0.8\linewidth]{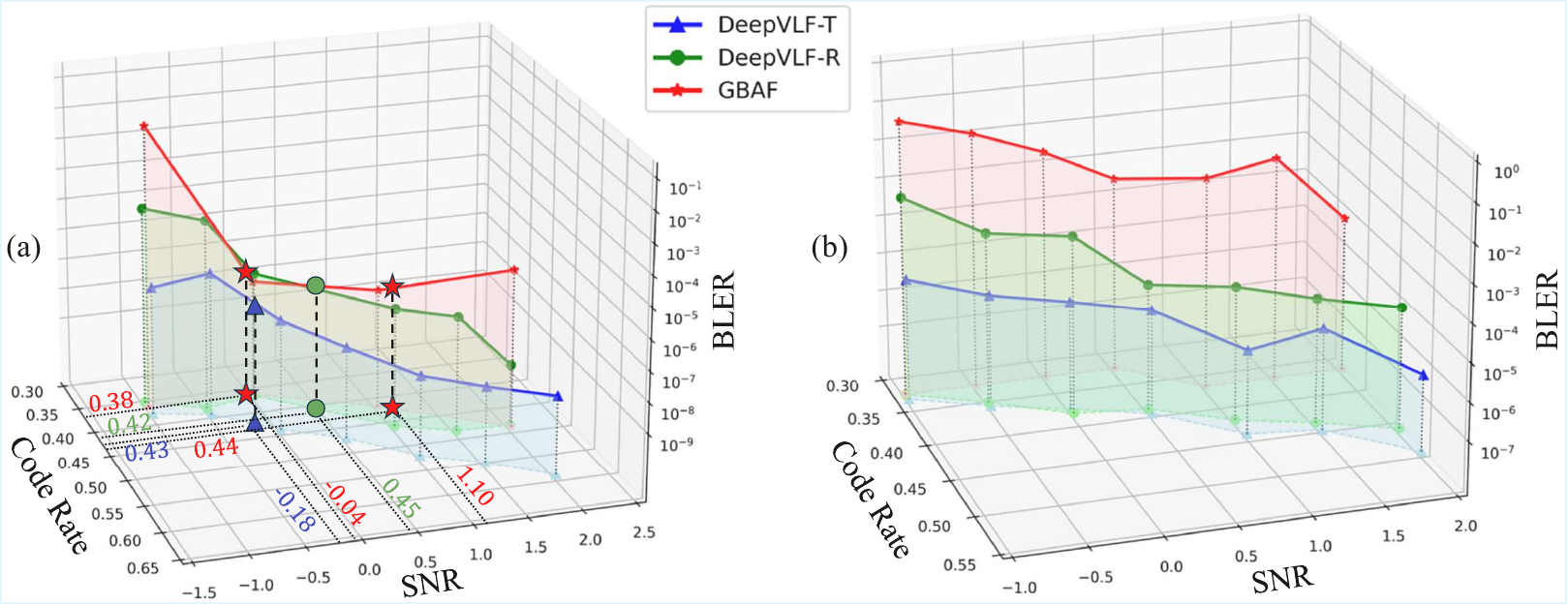}
    \caption{BLER as a joint function of forward SNR $\eta_f$ and average code rate $\mathbb{E}(R)$: (a) noiseless feedback; (b) $\eta_b =20 \text{dB}$.}
    \label{fig:awgn_3dplot}
\end{figure*}

\subsection{Performance over AWGN channels}

We first benchmark our DeepVLF codes against existing DL-based feedback codes over the classical AWGN channel. 
Fig.~\ref{fig:DeepVLF}(a) and \ref{fig:DeepVLF}(b) present the BLER versus code rate at forward SNRs of 1 dB and 0 dB, respectively. As shown, both DeepVLF-R and DeepVLF-T offer substantial performance gains over existing feedback codes, establishing a new state of the art, especially in the high-code-rate regimes. Consider the operation point at $\eta_f = 1$ dB (Fig.~\ref{fig:DeepVLF}(a)) and a target BLER of $10^{-5}$. 
\begin{itemize}[leftmargin=0.5cm]
    \item The GBAF code achieves this reliability at a code rate of approximately $0.458$ bits/channel use. In contrast, DeepVLF-R operates at $0.505$ bits/channel use for the same BLER, achieving a $10\%$ higher spectral efficiency. This translates to DeepVLF-R requiring only about $91\%$ of the channel uses that GBAF needs to deliver the same information with identical reliability.
    \item The advantage of transmitter-side control is further crystallized by DeepVLF-T. By leveraging its perfect knowledge of the message to make a global termination decision, it achieves remarkable ultra-reliability. At a code rate of $0.63$ bits/channel use, DeepVLF-T reaches a BLER around $10^{-9}$, effectively pushing the error floor several orders of magnitude lower than any fixed-length scheme.
\end{itemize}

The above patterns hold at the more challenging SNR of $0$ dB (Fig.~\ref{fig:DeepVLF}(b)), where both DeepVLF variants maintain a decisive advantage over GBAF across the entire rate range.

Fig.~\ref{fig:DeepVLF}(c) demonstrates that the gains are not predicated on ideal feedback. At $\eta_f = 1$ dB, DeepVLF-R, which strategically feeds back only minimal control information (decoded group indices) on a reliable side channel, maintains a robust performance curve that dominates GBAF. DeepVLF-T, which feeds back the full belief matrix, employs the robustness threshold $\widetilde{\gamma}=0.7$ to effectively filter out spurious termination triggers caused by feedback noise. Despite the noisy feedback link, it sustains a significant performance lead, particularly in the high-rate region where its global stopping rule is most valuable.

In addition to the fixed-SNR comparisons, Fig.~\ref{fig:awgn_3dplot} plots BLER as a joint function of forward SNR ($\eta_f$) and average code rate ($\mathbb{E}(R)$) as a three-dimensional surface, for both noiseless and noisy feedback.
In both figures, the DeepVLF performance surfaces lie consistently below and to the right of the GBAF surface in SNR-rate-BLER space. This means that, for a given SNR, the proposed schemes transmit the same information using fewer channel uses; for a given code rate, they achieve significantly lower error probabilities. These results show that DeepVLF extends the achievable operating region toward higher rates and stricter reliability levels, a regime that is largely inaccessible to fixed-length coding schemes like GBAF.

In Fig.~\ref{fig:awgn_3dplot}(a), we first trace the $10^{-7}$ BLER contour to quantify the resource efficiency gain. As shown, the baseline GBAF scheme achieves only a $0.38$ code rate at $-0.04$ dB SNR. Raising its rate to $0.44$ requires increasing the SNR to about $1.10$ dB. In contrast, DeepVLF-T attains a similar rate ($0.43$) at a lower SNR of $-0.18$ dB, yielding an SNR gain of roughly $1.3$ dB over GBAF at comparable rates. DeepVLF-R performs between the two, reaching a rate of about $0.42$ at $0.45$ dB.

Beyond these individual points, Fig.~\ref{fig:awgn_3dplot}(a) shows that for SNRs between 0 and 2 dB, the average code rate of DeepVLF-T is approximately $20$-$25\%$ higher than that of GBAF, while that of DeepVLF-R is $1$-$10\%$ higher. This advantage grows especially pronounced in the high-rate regime: while GBAF suffers a rapid rise in BLER, DeepVLF maintains reliability by adaptively terminating transmissions.

Fig.~\ref{fig:awgn_3dplot}(b) summarizes performance under noisy feedback. Although feedback noise degrades all schemes, the minimum BLER of GBAF remains above the maximum BLER of DeepVLF-T across the entire SNR range. For forward SNRs between $0$ and $2$ dB, DeepVLF-T provides code-rate gains of about $15$-$33\%$ relative to GBAF, while DeepVLF-R achieves gains of $14$-$24\%$. Moreover, average BLER is reduced by roughly two orders of magnitude for DeepVLF-T and nearly three orders for DeepVLF-R compared with GBAF.

\subsection{Performance Evaluation over 5G-NR Fading Channels}

To rigorously assess the viability of DeepVLF, we evaluate its performance under a 3GPP TR 38.901 NR CDL-C fading channel \cite{3gppTR38901}, moving beyond the controlled AWGN assumption.
We model the uplink communication between a mobile user (node A/UE) and a base station (node B/gNB). The CDL-C model for non-line-of-sight (NLOS) propagation is implemented using the Quasi Deterministic Radio Channel Generator (QuaDRiGa) \cite{jaeckel2014quadriga}. The system operates at a carrier frequency of $3.5$ GHz with a subcarrier spacing of $30$ kHz, and the user moves at a speed of $v_u=1$ m/s with an RMS delay spread of $100$ ns.

Using QuaDRiGa, we generate two independent long-scale fading trajectories, $\text{Tr}_t(v_u)$ and $\text{Tr}_e(v_u)$, as continuous ``paths'' of channel realizations in the training and evaluation phases, respectively.
During training, for each batch, a random starting point is selected in $\text{Tr}_t(v_u)$, and the channel gains for the subsequent $\tau_{\text{max}}$ consecutive time slots (each slot corresponds to one communication round) are extracted to form the fading coefficients for that training sample. This ensures the model is exposed to a wide variety of channel dynamics. The evaluation phase follows the same procedure using the held-out trajectory $\text{Tr}_e(v_u)$ to guarantee an unbiased test.

The $Q$ real-valued parity symbols in each round are modulated onto $\lceil Q/2\rceil$ subcarriers. We assume perfect channel state information (CSI) at the receiver (gNB). Therefore, the effect of fading can be compensated by applying the inverse channel gains, transforming the system model to:
\begin{equation*}
    \bm{y}^{(\tau)}\odot\frac{1}{\bm{h}^{(\tau)}}=\bm{x}^{(\tau)}+\bm{w}^{(\tau)}\odot\frac{1}{\bm{h}^{(\tau)}},
\end{equation*}
where $\odot$ denotes the element-wise product; $\bm{h}^{(\tau)}$ is the vector of complex channel gains. The effective noise $\bm{w}^{(\tau)}\odot\frac{1}{\bm{h}^{(\tau)}}$ is now non-Gaussian and channel-dependent. 

\begin{figure}[t]
    \centering
    \includegraphics[trim={9cm 4.5cm 7.5cm 4cm},clip,width=0.8\linewidth]{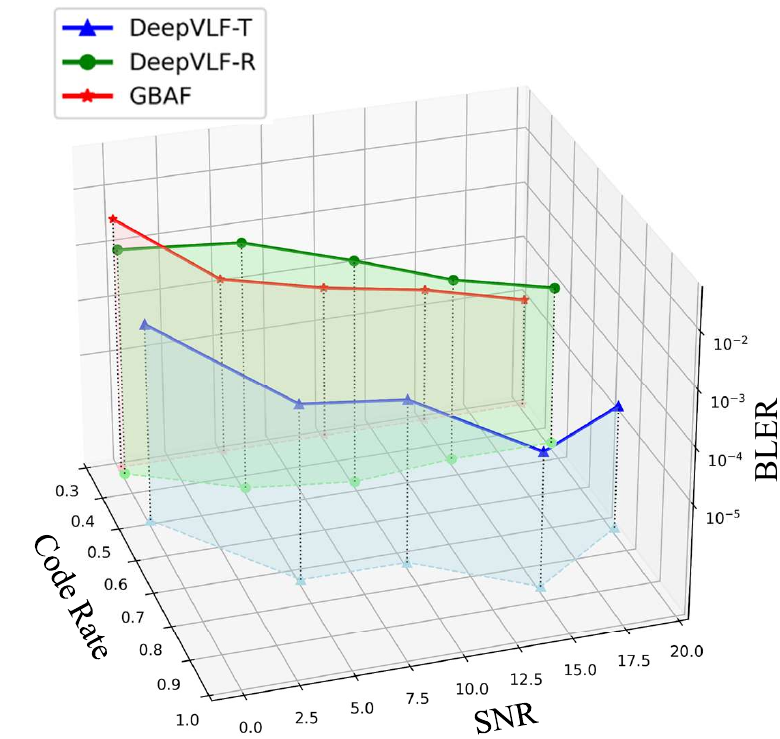}
    \caption{Performance comparison of DeepVLF-R, DeepVLF-T, and the GBAF code over 5G-NR fading channels with noiseless feedback.}
    \label{fig:fading}
\end{figure}

Fig.~\ref{fig:fading} presents the core performance comparison over the fading channel with noiseless feedback, plotting BLER as a function of forward SNR and average code rate. The results demonstrate that the advantages of DeepVLF extend robustly to this dynamic propagation environment. Specifically,
\begin{itemize}[leftmargin=0.5cm]
    \item DeepVLF-R consistently achieves a higher average code rate than the fixed-length GBAF baseline for any given SNR and BLER target. For instance, at $\eta_f=0$ dB and a BLER of $10^{-3}$, DeepVLF-R operates at a code rate approximately $7\%$ higher. This gain stems from its ability to opportunistically terminate transmission for individual bit groups during favorable channel realizations.
    \item DeepVLF-T exhibits even more pronounced gains in spectral efficiency. Its transmitter-side global verification allows it to achieve a target BLER with significantly fewer channel uses. Quantitatively, DeepVLF-T requires up to $55\%$ fewer channel uses on average than GBAF to deliver the same number of information bits at comparable reliability (e.g., BLER $\sim 10^{-4}$ at $\eta_f=5$ dB).
\end{itemize}

The sustained superiority confirms that both DeepVLF variants can learn a robust adaptive coding strategy that effectively combats the multiplicative and time-varying distortion introduced by fading.

\subsection{Analysis of Learned Encoding Dynamics}

\begin{figure}[t]
    \centering
    \includegraphics[width=1\linewidth]{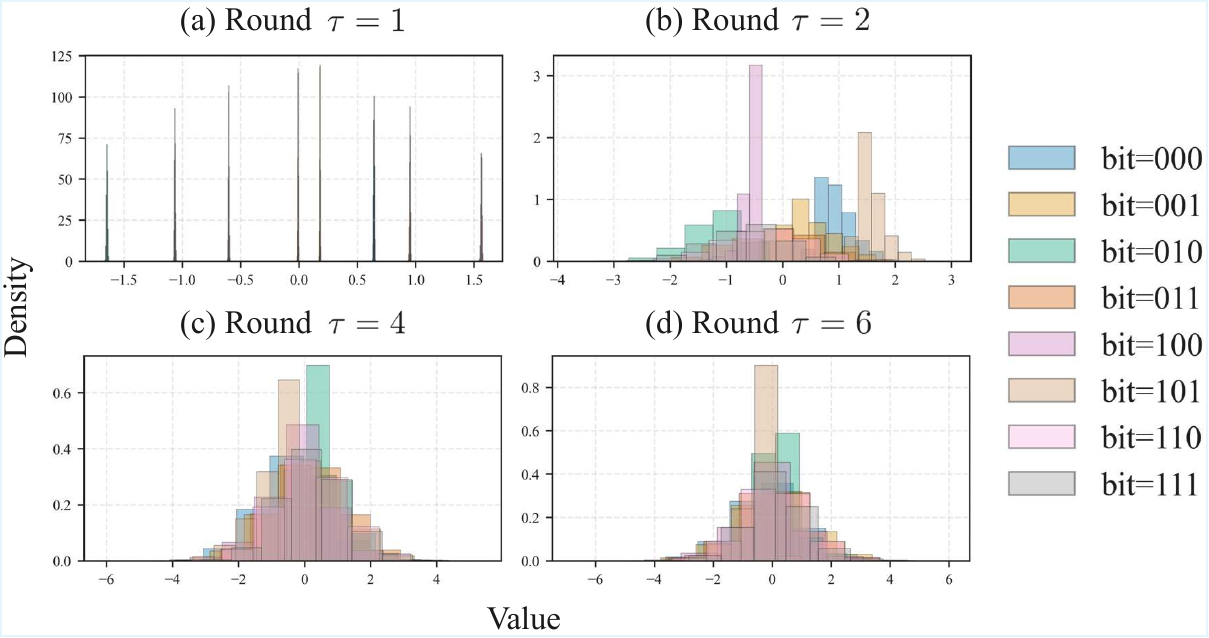}
    \caption{Evolution of transmitted symbol distribution for a targeted bit group under controlled conditions. Kernel density estimates show the distribution of the first parity symbol $x_1^{(\tau)}$ associated with bit group $\bm{b}_1$, across communication rounds $\tau$. Each color represents one of the 8 possible patterns of $\bm{b}_1$.}
    \label{fig:first_parity}
\end{figure}

To move beyond empirical performance metrics and probe the internal strategies learned by DeepVLF, we conduct a controlled visualization experiment. The objective is to isolate and understand how the encoder allocates information across successive communication rounds, specifically, whether it spontaneously discovers principles akin to classical feedback coding schemes.

We design a setup to trace the causal relationship between a specific information bit group and its corresponding transmitted parity symbol. For a single communication session (i.e., one transmitted block), we hold constant all sources of randomness and variation except one:
\begin{itemize}[leftmargin=0.5cm]
    \item The channel noise vectors $\bm{w}$ and $\bm{\widetilde{w}}$ are fixed.
    \item The information bits for groups $\bm{b}_2,...,\bm{b}_Q$ are fixed.
    \item Only the first bit group, $\bm{b}_1$, is varied across all $2^m=8$ possible patterns.
\end{itemize}

We then record the value of the first transmitted parity symbol, $x_1^{(\tau)}$, generated for the first check node (which is directly associated with $\bm{b}_1$) in each communication round. To ensure statistical reliability and generalizability across different noise and message backgrounds, this entire procedure is repeated over $800,000$ independent trials. In each trial, new realizations of the fixed noise and the constant bit groups $\bm{b}_2,...,\bm{b}_Q$ are randomly generated. This massive repetition ensures the observed distributions reflect the true, averaged behavior of the encoder.

Fig.~\ref{fig:first_parity} presents the resulting kernel density estimates of the $x_1^{(\tau)}$ distributions for each of the 8 possible $\bm{b}_1$ patterns, across rounds $\tau=1,2,4,6$.
\begin{itemize}[leftmargin=0.5cm]
    \item \textbf{Round 1 (message-centric phase)}: In the first transmission, the distributions of $x_1^{(\tau)}$ for different $\bm{b}_1$ patterns are well-separated into distinct clusters. This demonstrates that the encoder primarily invests the initial channel use to send a discriminative, message-dependent signal. The parity symbol is highly informative of the specific bit group value.
    \item \textbf{Subsequent rounds (noise-centric refinement phase)}: From Round 2 onward, the distributions begin to overlap significantly. By Rounds $4$ and $6$, the eight distributions converge almost completely. This indicates a dramatic shift in strategy. In later rounds, the transmitted symbol $x_1^{(\tau)}$ carries minimal new message information about $\bm{b}_1$. Instead, its value appears to be optimized to help the decoder refine its estimate by effectively canceling the accumulated noise from previous forward transmissions, leveraging the feedback link.
\end{itemize}

The learned two-phase strategy, where the first transmission is maximally informative of the message, and subsequent transmissions are dedicated to noise estimation and cancellation, is the fundamental principle underlying the celebrated Schalkwijk-Kailath (SK) scheme \cite{schalkwijk1966coding1}. Remarkably, this sophisticated coding strategy is not architected into DeepVLF via constraints or regularization. It emerges autonomously through end-to-end training to minimize the combined loss function under the power constraint. This finding provides a critical interpretability bridge: it demonstrates that DL-based feedback codes can inherently discover coding strategies that are provably optimal in an information-theoretic sense, validating the functional efficacy of the data-driven approach and aligning learned models with established theoretical principles.

\section{Conclusion}\label{sec:conclusion}
This work has introduced a foundational framework for variable-length feedback coding powered by deep learning, demonstrating that adaptive communication strategies can be effectively learned from data rather than solely derived from analytical design. By moving beyond the fixed-block-length paradigm, DeepVLF achieves substantial gains in spectral efficiency and reliability, particularly in the high-rate regime where conventional learned codes falter. More importantly, the framework reveals a principled convergence between data-driven methods and classical information theory, affirming that well-structured learning can inherently align with information-theoretically sound principles.

The implications of this research extend beyond immediate performance improvements. It underscores that variable-length coding is not merely an optimization tool but a conceptual shift toward autonomous, context-aware communication. Looking ahead, this approach invites further integration with semantic and goal-oriented feedback, promising to enable a new generation of networks that are not only faster and more reliable, but also inherently adaptive and efficient.

\bibliographystyle{IEEEtran}
\bibliography{Ref}

\end{document}